\documentclass[final]{aipproc}
\layoutstyle{6x9}
\begin{document}
\title{Chiral properties of the constituent quark model}
\author{Wolfgang Lucha}{address={HEPHY, Austrian Academy of Sciences,
Nikolsdorfergasse 18, A-1050 Vienna, Austria}}\iftrue
\author{Dmitri Melikhov$^{*,}$}{address={Institute of Nuclear Physics,
Moscow State University, 119992, Moscow, Russia }}
\author{Silvano Simula }{address={INFN, Sezione di Roma 3, Via della
Vasca Navale 84, I-00146, Roma, Italy}}\fi

\begin{abstract}
We show that, in a model based exclusively on
constituent-quark degrees of freedom interacting via a potential,
the full axial current is conserved {\em if\/} the spectrum of
$\bar QQ$ states contains a massless pseudoscalar. The current
conservation emerges nonperturbatively if the model satisfies
certain constraints on (i) the axial coupling $g_A$ of the
constituent quark and (ii) the $\bar QQ$ potential at large
distances. We define the chiral point of the constituent quark
model as that set of values of the parameters (such as the masses
of the constituent quarks and the couplings in the $\bar QQ$
potential) for which the mass of the lowest pseudoscalar $\bar QQ$
bound state vanishes. At the chiral point the main signatures of
the spontaneously broken chiral symmetry are shown to be present,
namely: the axial current is conserved, the decay constants of the
excited pseudoscalar bound states vanish, and the pion decay
constant has a nonzero value.
\end{abstract}
\maketitle

%\section{Constituent-quark axial current conservation}
Chiral symmetry
is a basic symmetry of massless QCD which, apart from the axial
anomaly in the flavor-singlet channel, entails the conservation of
the axial-vector current. The masses of the light {\sl u\/} and
{\sl d\/} quarks are small compared to the confinement scale, and
consequently the chiral limit serves as a good approximation for
the light-quark sector of QCD. Chiral symmetry in QCD is
spontaneously broken, and is thus not a symmetry~of the hadron
spectrum: except for the existence of the octet of light
pseudoscalar mesons, the lowest-energy part of the hadron spectrum
shows no trace of chiral symmetry.

Because of confinement, the calculation of the hadron mass
spectrum directly from~the QCD Lagrangian is a very challenging
task, which requires a nonperturbative approach. QCD-inspired
constituent quark models (i.e., models based on constituent-quark
degrees of freedom in which mesons appear as $\bar QQ$ bound
states in a potential) proved to be quite successful for the
description of the mass spectrum of hadrons and their interactions
at low momentum transfers \cite{gromes,gi,melikhov}. Because of
the proper description of the hadron mass spectrum, the Lagrangian
of the constituent quark model cannot be chirally invariant: it
would produce a chirally invariant spectrum of hadron states.
Consequently, the Noether axial current found in such models is
not conserved but satisfies the divergence equation
\begin{equation}
\label{divQ}
\partial^\mu[\bar Q(x)\gamma_\mu\gamma_5Q(x)]=2m_Q\bar Q(x){\rm
i}\gamma_5Q(x).
\end{equation}
In a recent paper \cite{lms} we have shown that,
nevertheless, taking into account the infinite number of diagrams
describing the $\bar QQ$ interactions, leads to the full axial
current of the constituent quarks, which turns out to have the
structure
%\begin{equation}
%\label{Eq:CQAC}
%\langle0|\bar
%q\gamma_\mu\gamma_5q|\bar QQ\rangle=g_A(p^2)\left(\bar
%Q\gamma_\mu\gamma_5Q+2m_Q\frac{p_\mu}{p^2}\bar
%Q\gamma_5Q+2m_Q\frac{p_\mu}{p^2}\bar
%Q\gamma_5Q\,\frac{O(M_\pi^2)}{p^2-M_\pi^2}\right).
%\end{equation}
\begin{eqnarray}
\label{Eq:CQAC}
\langle 0|\bar q \gamma_\mu \gamma_5 q |\bar QQ\rangle=\hspace{-.1cm}
g_A(p^2)\hspace{-.1cm} \left\{\bar Q \gamma_\mu \gamma_5 Q
+2m_Q \frac{p_\mu}{p^2}\bar Q \gamma_5 Q\right\}\hspace{-.1cm}
+g_A(p^2)\frac{p_\mu}{p^2}\bar Q \gamma_5 Q\,
\frac{2m_Q O(M_\pi^2)}{p^2-M_\pi^2}.
\end{eqnarray}
Obviously, the term in curly brackets is transverse by virtue of
Eq.~({\ref{divQ}). Therefore, the full ``constituent-quark axial
current'' is conserved if the mass $M_\pi$ of the pion, the lowest
$\bar QQ$ pseudoscalar bound state, vanishes. As has been
demonstrated in Ref.~\cite{lms}, to guarantee the axial current
conservation up to terms of order $O(M_\pi^2)$ requires that the
axial coupling $g_A$ of the constituent quarks is not constant but
that it is related to the pion wave function $\Psi_\pi(s)$ by
\begin{equation}
\label{3}
g_A(s)=\eta_A(s-M_\pi^2)\Psi_\pi(s)+O(M_\pi^2),\quad\eta_A=
\mbox{const}.
\end{equation}
It should be recalled here that the
spontaneous breaking of chiral symmetry requires not only the
conservation of the axial current, but also the nonvanishing of
the coupling of a massive fermion (such as a nucleon or a
constituent quark) to the pion, i.e., $g_A(s)$ should be nonzero
at $s=M_\pi^2.$ Consequently, to be compatible with the
spontaneous breaking of chiral symmetry, the potential model
should generate a light pseudoscalar bound~state~for which
$\Psi_\pi(s=M_\pi^2)$ has a pole at $s=M_\pi^2$ \cite{lms}.

In Ref.~\cite{lms} it is shown that the behavior of $\Psi_\pi(s)$
at $s=M_\pi^2$ is related to the behavior~of the potential of the
$\bar QQ$ interaction at large separations $r.$ More precisely,
$\Psi_\pi(s)$ exhibits~a pole at $s=M_\pi^2$ only if the potential
saturates at large $r$:
\begin{equation}
V(r\to\infty)={\rm const}<\infty.
\end{equation}
In
this case the nearly massless pion is a strongly bound $\bar QQ$
state with binding energy $\epsilon\simeq 2m.$

The observed conservation of the axial current allows us to define
the {\em chiral point~of the constituent quark model\/} as exactly
that set of values of the parameters which leads to a massless
lowest pseudoscalar $\bar QQ$ bound state.\footnote{Precisely, we
consider a given potential (i.e., with fixed couplings) and adjust
the constituent mass of~the light quark until $M_\pi$ vanishes.}
In the following, let us consider~certain properties of the
constituent quark model at the chiral point.

\vspace{-.4cm}
\subsubsection{Decay constants of pseudoscalar mesons}
\vspace{-.2cm}

Making use of the
relation (\ref{3}) between the axial coupling of the constituent
quark, $g_A(s),$ and the pion wave function, the standard
quark-model expression for the decay constant of the $n$-th
excitation of a pseudoscalar meson \cite{melikhov} takes the form
\cite{lms}
\begin{equation}
\label{Eq:fP}
f_{P}(n)=2m_Q\eta_A\sqrt{N_c}\int{\rm d}s\,\Psi_{0}(s)
\Psi_{n}(s)\rho(s,m_Q^2,m_Q^2)\frac{s-M_\pi^2}{s}+O(M_\pi^2).
\end{equation}
The wave functions $\Psi_{n}(s)$ of the pseudoscalar
states satisfy the orthogonality
condition
\begin{equation}
\label{orthonorm}
\int{\rm d}s\,\Psi_{n}(s)\Psi_{m}(s)\rho(s,m_Q^2,m_Q^2)=\delta_{mn}.
\end{equation}
For the ground state, $n=0,$ the decay constant
$f_{P}(0)\equiv f_\pi$ is clearly finite in the chiral limit. With
the help of Eq.~(\ref{orthonorm}), we obtain the relation
\cite{lms}
\begin{equation}
f_\pi=2m_Q\eta_A\sqrt{N_c}+O(M_\pi^2).
\end{equation}
For excited
states, $n\ne 0,$ Eq.~(\ref{Eq:fP}) implies \cite{lms}, by virtue
of the orthogonality condition~(\ref{orthonorm}),
\begin{equation}
f_{P}(n\ne0)=-2m_Q\eta_AM_\pi^2\int{\rm
d}s\,\Psi_{0}(s)\Psi_{n}(s)\frac{\rho(s,m_Q^2,m_Q^2)}{s}+O(M_\pi^2).
\end{equation}
This decay constant is proportional to $M_\pi^2$ and therefore
vanishes in the chiral limit,~in accordance with the equations of
motion in QCD. Also, beyond the chiral limit the~decay constants
of the excited pseudoscalars are expected to be strongly
suppressed compared to the pionic decay constant $f_\pi$
\cite{lucha}. However, all more accurate predictions for the decay
constants of the excited pseudoscalars require a better knowledge
of the details of $g_A(s),$ since in this case the unknown terms
of the order $O(M_\pi^2)$ are of the same order as the
contribution given by the main term in $g_A(s).$

\vspace{-.4cm}
\subsubsection{Pionic coupling of hadrons}
\vspace{-.2cm}

The result (\ref{Eq:CQAC}) for the full axial current contains an
explicit pion pole, thus providing the possibility to extract the
amplitude $A(h_1\to h_2\pi)$ for pionic decays $h_1\to h_2+\pi$
\cite{lms}:
\begin{equation}
\label{amp}
p_\mu A(h_1\to
h_2\pi)=\lim_{p^2\to M_\pi^2}\frac{p^2-M_\pi^2}{f_\pi}\langle
h_2|j^5_\mu|h_1\rangle =p_\mu \frac{2m_Q}{f_\pi}\langle h_2|\bar
Q\gamma_5 Q|h_1\rangle.
\end{equation}
It is understood that the
amplitude $\langle h_2|\bar Q\gamma_5 Q|h_1\rangle$ is calculated
in terms of the constituent quark description of the hadrons $h_1$
and $h_2.$ The expression (\ref{amp}) for the amplitude has~been
successfully applied to pionic decays of charmed mesons
\cite{charm}.

\vspace{-.4cm}
\subsubsection{\it The chiral constituent quark mass}
\vspace{-.2cm}

Clearly, the constituent quark mass does not vanish in the chiral
point. We give now an estimate for the constituent quark mass
corresponding to the chiral limit, $m_Q^0$, making use of the
following relation between the constituent quark mass $m_Q$ and
the current quark mass $m$ at the chiral-symmetry breaking scale
$\mu_\chi\simeq 1$ GeV \cite{ms2004}:
\begin{eqnarray}
\label{ms}
\langle\bar qq\rangle=\frac{N_c}{\pi^2}\int_0^\infty
dk\,k^2\exp(-k^2/\beta_\infty^2)\left\{\frac{m}{\sqrt{m^2+k^2}}-
\frac{m_Q}{\sqrt{m_Q^2+k^2}}\right\},
\end{eqnarray}
with $\beta_\infty\simeq 0.7$ GeV \cite{ms2004}. We now have to
take into account the dependence of the quark condensate on the
value of the current quark mass. For the physical value of the
quark condensate, corresponding to the current quark mass $m=6$
MeV, we use $\langle \bar qq\rangle=-(240\pm 15\,\mbox{ MeV})^3$.
Eq.~(\ref{ms}) then gives $m_Q=220$ MeV, a typical value of the
$u$ and $d$ constituent quark mass \cite{gi}. In order to consider
the chiral limit, $m\to 0$, the dependence of the quark condensate
on the current quark mass should be taken into account. Setting
$m=0$, and making use of the chiral quark condensate $\langle \bar
qq\rangle_{m=0}\simeq -(230\pm 15\,\mbox{ MeV})^3$, Eq.~(\ref{ms})
gives the chiral constituent quark mass $m_Q^0=180$ MeV. Let us
notice that this is precisely the value of the chiral constituent
quark mass of the Godfrey--Isgur model~\cite{gi}.

\newpage
In summary, we have demonstrated that the relativistic quark
picture based exclusively on constituent-quark degrees of freedom
is fully compatible with the (well-known)~chiral properties of QCD
{\em if\/} it encompasses the following
features:
\begin{itemize}
\item
The axial coupling $g_A$ of the
constituent quarks is a momentum-dependent quantity, $g_A=g_A(s),$
and is related to the pion $\bar QQ$ wave function.
\item
The $\bar
QQ$ potential $V(r)$ saturates at large interquark separations:
$V(r\to\infty)\to\mbox{const}.$
\end{itemize}
Under the above
conditions, a summation of the infinite number of diagrams
describing constituent-quark soft interactions leads to the full
axial current of the constituent~quarks which is then conserved up
to terms of order $O(M_\pi^2).$

We defined the chiral point of the constituent quark model as that
set of values of the parameters of the model (masses of the
constituent quarks and couplings in the quark potential) for which
the mass of the lowest pseudoscalar $\bar QQ$ bound state,
$M_\pi,$ vanishes. Although the constituent quark mass clearly
does not vanish at the chiral point, we claim that the chiral
point of the constituent quark model corresponds to the
spontaneously broken chiral limit of QCD for the following three
reasons. (i) At the chiral point the~full nonperturbative axial
current of the constituent quarks is conserved (without the
explicit introduction of Goldstone degrees of freedom). (ii) The
lowest-energy part of the hadron spectrum has no other traces of
chiral symmetry except for a massless pseudoscalar. (iii) Two
important signatures of the spontaneously-broken chiral symmetry
can be seen: the decay constant $f_\pi$ of the massless pion is
finite, that means, nonvanishing, whereas~all~the decay constants
of the excited massive pseudoscalars vanish.

We emphasize that the nonperturbative emergence of chiral symmetry
in a model with merely constituent-quark degrees of freedom
\cite{lms} is qualitatively different from the chiral symmetry of
models which explicitly contain Goldstones along with constituent
quarks: the latter may be rendered chirally invariant for any
value of the constituent quark mass, whereas in our approach
chirally symmetry is present only for a definite (nonvanishing)
value of the constituent quark mass which leads to a massless
ground-state pseudoscalar.

%\begin{theacknowledgments}

\vspace{.4cm}

{\it Acknowledgments.} D.~M.~was supported by the Austrian
Science Fund (FWF) under project No.~P17692.
%\end{theacknowledgments}


\begin{thebibliography}{99}
\bibitem{gromes}W.~Lucha, F.~F.~Sch\"oberl, and D.~Gromes,
\emph{Phys.~Rep.}~\textbf{200}, 127 (1991).
\bibitem{gi}S.~Godfrey and N.~Isgur, \emph{Phys.~Rev.~D} \textbf{32},
189 (1985).
\bibitem{melikhov}V.~V.~Anisovich {\em et al.}, \emph{Nucl.~Phys.~A}
\textbf{544}, 747 (1992); F.~Cardarelli {\em et al.},
\emph{Phys.~Lett.~B} \textbf{332}, 1 (1994); D.~Melikhov,
\emph{Phys.~Rev.~D} \textbf{53}, 2460 (1996); F.~Cardarelli {\em
et al.}, \emph{Phys.~Rev.~D} \textbf{53}, 6682 (1996); D.~Melikhov
and B.~Stech, \emph{Phys.~Rev.~D} \textbf{62}, 014006 (2000).
\bibitem{lms}W.~Lucha, D.~Melikhov, and S.~Simula, \emph{Phys.~Rev.~D}
\textbf{74}, 054004 (2006).
\bibitem{lucha}M.~A.~Shifman, A.~I.~Vainshtein, and V.~I.~Zakharov,
\emph{Nucl.~Phys.~B} \textbf{147}, 385 (1979); A.~H\"oll,
A.~Krassnigg, and C.~D.~Roberts, \emph{Phys.~Rev.~C} \textbf{70},
042203(R) (2004); W.~Lucha and D.~Melikhov, \emph{Phys.~Rev.~D}
\textbf{73}, 054009 (2006).
\bibitem{charm}D.~Melikhov and O.~Pene, \emph{Phys.~Lett.~B}
\textbf{446}, 336 (1999); D.~Melikhov and M.~Beyer,
\emph{Phys.~Lett.~B} \textbf{452}, 121 (1999); D.~Melikhov and
B.~Stech, \emph{Phys.~Rev.~D} \textbf{74}, 034022 (2006).
\bibitem{ms2004}D.~Melikhov and S.~Simula, \emph{Eur.~Phys.~J.~C}
\textbf{37}, 437 (2004).\end{thebibliography}
\end{document}